\documentclass[11pt,reqno,oneside]{amsart}
\usepackage{amssymb}
\usepackage[reqno]{amsmath}

\begin{document}

\begin{flushright}
\baselineskip=12pt
CERN--TH/97--168\\
\tt gr-qc/9707039
\end{flushright}

\title[Canonical General Relativity]{Canonical General Relativity: the Diffeomorphism constraints and Spatial Frame Transformations} 
\author{M. A. Clayton}
\address{CERN--Theory Division, CH--1211 Geneva 23, Switzerland}
\email{Michael.A.Clayton@cern.ch}
\date{\today}

\begin{abstract}
Einstein's general relativity with both metric and vielbein treated as independent fields is considered, demonstrating the existence of a consistent variational principle and deriving a Hamiltonian formalism that treats the spatial metric and spatial vielbein as canonical coordinates.
This results in a Hamiltonian in standard form that consists of Hamiltonian and momentum constraints as well as constraints that generate spatial frame transformations---all appearing as primary, first class constraints on phase space.
The formalism encompasses the standard coordinate frame and vielbein approaches to general relativity, and the constraint algebra derived herein reproduces known results in either limit.
\end{abstract}
\maketitle

\section{Introduction}
\label{sect:intro}

A recent review of actions for general relativity (GR)~\cite{Peldan:1994} describes formulations in coordinate and (orthogonal) vielbein frames from variations of Einstein--Hilbert and Einstein--Palatini actions.
Notably absent is an action that encompasses both of these cases.
Such an action possesses the full general linear group invariance, as well as the usual diffeomorphism invariance, allowing the ``limit'' to orthogonal vielbein and coordinate frame approaches as different choices of gauge.
This structure is represented on the full arbitrariness of the choice of vielbein ($16$ functions) as well as those of the spacetime metric ($10$ functions).
That one can construct such an action is perhaps no surprise~\cite{Hehl+:1995}, however we will also construct a Hamiltonian for the system in which the spatial $\mathrm{GL}(3,\mathbb{R})$ invariance is represented by infinitesimal generators on phase space.
The resulting constraint algebra is derived, complete with conditions that determine surface contributions that guarantee that the field equations of GR follow properly from the Hamiltonian~\cite{Regge+Teitelboim:1974}.

As is always the case in such constructions, the Hamiltonian is far from unique.
We will instead be focusing on the combined algebra of infinitesimal diffeomorphisms and $\mathrm{GL}(3,\mathbb{R})$ transformations which is uniquely determined once the parameterization of phase space and the action of the generators is fixed.
In an earlier work~\cite{Clayton:1996b} this algebra was derived through a generalization of the geometric argument of Teitelboim~\cite{Teitelboim:1973,Teitelboim:1980}, which led to a ``derivation'' of canonical GR~\cite{Hojman+Kuchar+Teitelboim:1976}.
This generalized algebra has appeared previously in the examination of (orthonormal) tetrad GR~\cite{Henneaux:1983,Charap+Henneaux+Nelson:1988} and here we extend this type of analysis to the more general case, including an arbitrary tetrad as well as metric degrees of freedom.

It is noteworthy that we are \textit{not} attempting to implement the full set of $\mathrm{GL}(4,\mathbb{R})$ generators~\cite{Charap+Henneaux+Nelson:1988,Floreanini+Percacci:1990}, nor the full set of spacetime diffeomorphisms~\cite{Isham+Kuchar:1985a,Isham+Kuchar:1985b}---instead we view the problem as purely geometrodynamic.
Initial data consisting of the components of the spatial metric and vielbein that satisfies the constraints are given on the initial hypersurface, we determine the infinitesimal transformations of this data that lead to an equivalent physical problem (frame rotations and spatial diffeomorphisms), and then determine the conditions on the generator of time evolution that guarantee that the evolution of the system is consistent with spacetime diffeomorphism invariance.

Inevitably we end up with a system with a higher degree of redundancy ($30$ phase space degrees of freedom and $13$ primary, first class constraints) and the possibility of new choices of gauge that are useful for numerical relativity exists; this will not be considered in this work.
Ultimately we are interested in exploring the relationship between diffeomorphism invariance, the strong equivalence principle, and the evolution of quantum systems.
In particular, whether it is possible to specify some type of quantum evolution that respects diffeomorphism invariance without the need to introduce an (approximate) timelike killing vector as is usually necessary~\cite{DeWitt:1975,Kuchar:1988}.
We would also like to quantify the non-invariance of the conventional curved spacetime field theory, towards an understanding of which results may survive in a more fundamental theory based on the strong equivalence principle, and which depend on a choice of frame.

\section{The General Frame Construction}
\label{sect:gen}

The natural setting for this work is the space of moving (vielbein) frames above a manifold $\mathbf{M}$, that is, the space of smooth assignments of a frame of reference above each point, equivalent to smooth sections of the general linear frame bundle $GL\mathbf{M}$.
Physical fields erected above $\mathbf{M}$ (the metric, curvature, matter fields, \textit{etc}\ldots) are then associated to this bundle through some representation of $\mathrm{GL}(n,\mathbb{R})$, and components may be defined in terms of said frame.
Standard presentations of classical general relativity often adopt coordinate (or holonomic) frames from the outset, introducing moving (or vielbein) frames as a `generalized' concept in Riemannian geometry.
Even when such frames are given equal footing with holonomic frames, in the variational principle, coordinate components are once again given precedence.
In order to demonstrate that this is unnecessary, we proceed to review some results from Riemannian geometry in a general linear frame.

\subsection{Moving Frames}

A moving frame $\{e_A\}$ above a point in $\mathbf{M}$ will be written in terms of a particular coordinate frame as $e_A:={E^\mu}_A\partial_\mu$ and the coframe that is dual to it as $\theta^A:={E^A}_\mu dx^\mu$, where the duality relation $\theta^A[e_B]=\delta^A_B$ implies that the vielbeins satisfy ${E^\mu}_A{E^A}_\nu=\delta^\mu_\nu$ and ${E^\mu}_B{E^A}_\mu=\delta^A_B$.
In this section we consider frames that are chosen to be a smooth section of the general linear frame bundle $GL\mathbf{M}$, however in Section~\ref{sect:Ham} we will consider spatial frames defined by $e_a:={E^i}_a\partial_i$, where $\partial_i$ are the partial derivatives with respect to coordinates on a spatial hypersurface.

The volume form in a linear frame is given by~\cite{Wald:1984}
\begin{equation}
\int_\mathbf{M} *\mathbf{1}
=\int_\mathbf{M}\sqrt{\lvert\mathrm{g}\rvert}
\theta^0\wedge\theta^1\wedge\theta^2\wedge\theta^3
=\int_{\mathbf{M}}d^4x\,\sqrt{\lvert\mathrm{g}\rvert}E,
\end{equation}
the last of which will be used here.
Using compatibility and vanishing torsion, Gauss' law takes the form
\begin{equation}\label{eq:Gauss}
\int_R dx\,E\sqrt{-\mathrm{g}}\nabla_A[V]^A
=\int_R dx\,\partial_\mu\bigl[{\underline{E}^\mu}_A\boldsymbol{V}^A\bigr]
=\int_{\partial R} dS_\mu\,{\underline{E}^\mu}_A\boldsymbol{V}^A
=\int_{\partial R} d\underline{S}_A\,\boldsymbol{V}^A,
\end{equation}
where the surface measure is defined in terms of the normal vector $n^A$ to the boundary $\partial R$ of the region $R\subset \mathbf{M}$ by $S_A:=*\mathbf{1}[n]_A$.
We have also made the definition $\underline{T}:=ET$ for tensors weighted by the vielbein density $E:=\det [{E^A}_\mu]$, and $\boldsymbol{T}:=\sqrt{\lvert\mathrm{g}\rvert}T$ for tensors weighted by the spatial metric density $\sqrt{\lvert\mathrm{g}\rvert}:=\sqrt{\lvert\det[\mathrm{g}_{AB}]\rvert}$.

Under the change of frame determined by ${M^A}_{B}\in\mathrm{GL}(n,\mathbb{R})$, the frame and coframe transform as $\theta^A\rightarrow\theta^B={M^B}_{A}\theta^A$, $e_A\rightarrow e_B =e_A{\lvert M^{-1}\rvert^A}_{B}$, where ${M^A}_{C}{\lvert M^{-1}\rvert^C}_{B} ={\lvert M^{-1}\rvert^A}_{C}{M^C}_{B} =\delta^A_B$, and similarly for the components of tensors.
From the infinitesimal forms ${M^A}_{B}=\delta^A_B +{\omega^A}_B$, and 
${\lvert M^{-1}\rvert^A}_{B} =\delta^A_B -{\omega^A}_B$, we define the generators ${\Delta^A}_{B}$ of $\mathfrak{gl}(n,\mathbb{R})$, acting, for example, on vectors, covectors and both types of density as
\begin{equation}\label{eq:infinit}
\Delta_{\tilde{\omega}} [T]_A=-{\omega^B}_AT_B,\quad
\Delta_{\tilde{\omega}} [T]^A={\omega^A}_BT^B,\quad
\Delta_{\tilde{\omega}}\bigl[\sqrt{\gamma}\bigr]
=-{\omega^A}_{A}\sqrt{\gamma},\quad
\Delta_{\tilde{\omega}}[E]={\omega^A}_{A} E,
\end{equation}
where $\tilde{\omega}$ represents the matrix ${\omega^A}_{B}\in\mathrm{GL}(n,\mathbb{R})$.
These generators satisfy the Lie algebra of $\mathfrak{gl}(n,\mathbb{R})$ given by the matrix commutator~\cite{Marsden+Ratiu:1994}
\begin{equation}\label{eq:lie left}
[\Delta_{\tilde{\omega}_1},\Delta_{\tilde{\omega}_2}]
=-\Delta_{\tilde{[\omega_1,\omega_2]}},
\end{equation}
where ${[\omega_1,\omega_2]^A}_{B}:={{\omega_1}^A}_{C}{{\omega_2}^C}_{B}-{{\omega_1}^C}_{B}{{\omega_2}^A}_{C}$.

Once one has a definition of parallel transport, the covariant derivative operator is defined.
We will assume that it acts on the components of tensors as, for example $\nabla_A[V]^B=e_A[V^B]+\Gamma^B_{AC}V^C$, and we may then define what it means for the components of the metric $\mathrm{g}_{AB}$ to be compatible with this connection:
\begin{equation}\label{eq:COMP}
\nabla_A[\mathrm{g}]_{BC}=e_A[\mathrm{g}_{BC}]-\mathrm{g}_{DC}\Gamma^D_{AB}-\mathrm{g}_{BD}\Gamma^D_{AC}=0.
\end{equation}
From the non--holonomic nature of the frame $\{e_A\}$ we have the structure constants~\cite{Nakahara:1990}
\begin{equation}\label{eq:ST C}
[e_A,e_B]={C_{AB}}^Ce_C,\quad 
{C_{AB}}^C={E^C}_\mu\bigl(e_A[{E^\mu}_B]-e_B[{E^\mu}_A]\bigr),
\end{equation}
and the vanishing of the torsion tensor $T(X,Y):=\nabla_X[Y]-\nabla_Y[X]-[X,Y]$ ($\nabla_X:=X^A\nabla_A$ and $[X,Y]:=\pounds_X[Y]$) results in the following relationship between $\Gamma^A_{[BC]}$ and the structure constants
\begin{equation}\label{eq:TORSION}
T^A_{BC}=2\Gamma^A_{[BC]}-{C_{BC}}^A=0.
\end{equation}
Throughout we denote (anti-)symmetrization by $[\,]$ and $(\,)$ respectively, \textit{i.e.}, $T_{[AB]}:=\tfrac{1}{2}(T_{AB}-T_{BA})$ and $T_{(AB)}:=\tfrac{1}{2}(T_{AB}+T_{BA})$.
This, combined with~\eqref{eq:COMP}, allows an explicit solution of $\Gamma^A_{BC}$ in terms of metric and vielbein components
\begin{multline}\label{eq:Gamma solution}
\Gamma^A_{BC}=\tfrac{1}{2}\mathrm{g}^{AD}
\bigl(e_B[\mathrm{g}_{CD}]+e_C[\mathrm{g}_{DB}]-e_D[\mathrm{g}_{BC}]\bigr)\\
+\tfrac{1}{2}\mathrm{g}^{AD}\bigl(\mathrm{g}_{BE}{C_{DC}}^E
+\mathrm{g}_{CE}{C_{DB}}^E\bigr)
+\tfrac{1}{2}{C_{BC}}^A,
\end{multline}
a combination of the standard coordinate frame and orthogonal frame results.
We will also require the contraction of the Riemann curvature tensor: 
${R^A}_{BCD}=e_C[\Gamma^A_{DB}]
-e_D[\Gamma^A_{CB}]+\Gamma^E_{DB}\Gamma^A_{CE}
-\Gamma^E_{CB}\Gamma^A_{DE}-{C_{CD}}^E\Gamma^A_{EB}$ to the Ricci tensor:
\begin{equation}
\label{eq:RICCI}
R_{AB}:=e_C[\Gamma^C_{BA}]-e_B[\Gamma^C_{CA}]
+\Gamma^D_{BA}\Gamma^C_{CD}-\Gamma^D_{CA}\Gamma^C_{BD}-{C_{CB}}^D\Gamma^C_{DA}.
\end{equation}

\subsection{The Einstein-Hilbert Action in a General Linear Frame}

A direct translation of the action for GR results in (we use $16\pi\mathrm{G}=\mathrm{c}=1$ throughout)
\begin{equation}\label{eq:SGR}
S_{\textsc{gr}}=-\int_\mathbf{M} d^{4}\!x\,E\sqrt{\lvert\mathrm{g}\rvert}
\mathrm{g}^{AB}R_{AB},
\end{equation}
where the spacetime is now described by a section of $\mathit{Riem}_\mathbf{M}\times \mathit{GL}\mathbf{M}$, namely, a Riemannian metric and non-degenerate linear frame above $\mathbf{M}$.

In a variation of the connection coefficients in $R_{AB}$ (\textit{i.e.}, not considering the variation of the structure constant in the definition~\eqref{eq:RICCI}) we find that $\delta_{\Gamma}R_{AB}=\nabla_C[\delta\Gamma]^C_{BA}
-\nabla_B[\delta\Gamma]^C_{CA}$, and therefore treating the components $\Gamma^A_{(BC)}$ as independent leads via a Palatini variation to the compatibility conditions~\eqref{eq:COMP}
\begin{equation}\label{eq:Palatini}
\frac{\delta S_{\textsc{gr}}}{\delta \Gamma^C_{(AB)}}
=-E\Bigl(\nabla_{e_C}[\mathbf{g}]^{BA}
-\nabla_{e_D}[\mathbf{g}]^{(AD}\delta^{B)}_C\Bigr)=0.
\end{equation}
Note that the action is \textit{not} set up to generate the torsion-free conditions~\eqref{eq:TORSION}, instead the variations $\delta\Gamma^A_{[BC]}$ must be determined in terms of the vielbein degrees of freedom using~\eqref{eq:TORSION} and the variation of~\eqref{eq:ST C}:
\begin{equation}\label{eq:delta c}
\delta{C_{AB}}^C
=\nabla_A\bigl[{E^C}_\mu\delta{E^\mu}_B\bigr]
-\nabla_B\bigl[{E^C}_\mu\delta{E^\mu}_A\bigr]
+\delta{E^\mu}_A{E^D}_\mu\Gamma_{DB}^C
-\delta{E^\mu}_B{E^D}_\mu\Gamma_{DA}^C.
\end{equation}
To perform the Einstein--Hilbert variation, requires the variation of~\eqref{eq:Gamma solution}
\begin{equation}\label{eq:delta gamma}
\begin{split}
\delta\Gamma^A_{BC}
=\delta{E^\mu}_B{E^D}_\mu\Gamma^A_{DC}
&+\tfrac{1}{2}\nabla_B\bigl[
\mathrm{g}^{AD}\delta\mathrm{g}_{CD}+{E^A}_\mu\delta{E^\mu}_C
-\mathrm{g}^{AD}\mathrm{g}_{CE}{E^E}_\mu\delta{E^\mu}_D\bigr]\\
&+\tfrac{1}{2}\nabla_C\bigl[
\mathrm{g}^{AD}\delta\mathrm{g}_{DB}-{E^A}_\mu\delta{E^\mu}_B
-\mathrm{g}^{AD}\mathrm{g}_{BE}{E^E}_\mu\delta{E^\mu}_D\bigr]\\
&-\tfrac{1}{2}\mathrm{g}^{AD}\nabla_D\bigl[
\delta\mathrm{g}_{BC}-\mathrm{g}_{BE}{E^E}_\mu\delta{E^\mu}_C
-\mathrm{g}_{CE}{E^E}_\mu\delta{E^\mu}_B\bigr],
\end{split}
\end{equation}
where we note that the presence of the first term, although implying that $\delta\Gamma^A_{BC}$ is no longer a tensor, is precisely what guarantees that the variation of a covariant derivative will remain covariant.

The variation of~\eqref{eq:SGR} will be performed by treating the densitized components of the inverse metric $\mathbf{g}^{AB}$ and the frame degrees of freedom ${\underline{E}^\mu}_A$ as independent variables, using
\begin{equation}\label{eq:delta_g R}
\delta\underline{\boldsymbol{R}}
=\underline{R}_{AB}\delta\mathbf{g}^{AB}
+\bigl[2{\boldsymbol{R}^A}_B{E^B}_\mu
-\tfrac{1}{n-1}\boldsymbol{R}{E^A}_\mu
\bigr]\delta{\underline{E}^\mu}_A
-E\nabla_A\nabla_B\bigl[\delta \boldsymbol{S}^{AB}\bigr],
\end{equation}
where $\delta\boldsymbol{S}^{AB}:=\delta\mathbf{g}^{AB}
+\mathrm{g}^{AB}\delta\sqrt{\lvert\mathrm{g}\rvert}
+2\mathbf{g}^{AC}{E^{B}}_\mu\delta{E^\mu}_C
-2\mathbf{g}^{AB}{E^C}_\mu\delta{E^\mu}_C$.
In computing this we have left the dimensionality $n$ of spacetime arbitrary and noted that $\delta\sqrt{-\mathrm{g}}=\mathrm{g}_{AB}\delta\mathbf{g}^{AB}/(n-2)$ and $\delta E={E^A}_\mu\delta{\underline{E}^\mu}_A/(n-1)$.
Discarding surface terms we find
\begin{equation}\label{eq:E-L}
\frac{\delta S_{\textsc{gr}}}{\delta \mathbf{g}^{AB}}=\underline{R}_{AB}=0,\quad
\frac{\delta S_{\textsc{gr}}}{\delta {\underline{E}^\mu}_A}
=2{\boldsymbol{R}^A}_B{E^B}_\mu-\tfrac{1}{n-1}\boldsymbol{R}{E^A}_\mu=0,
\end{equation}
which are Einstein's equations in empty spacetime.
The equivalence of the variational results from the metric and vielbein degrees of freedom is expected algebraically due to an argument by Floreanini and Percacci~\cite{Floreanini+Percacci:1990}, and  also since we know that Einstein's equations are covariant under frame transformations.
This variational principle is actually a specific case of the more general formalism that includes affine frames, non-metricity and torsion that appears in~\cite{Hehl+:1995}.

\section{Hamiltonian Formalism}
\label{sect:Ham}

We turn now to the construction of a Hamiltonian formalism for the system, much of which is a straightforward application of the Bergmann--Dirac procedure for constrained systems described in detail in~\cite{Isenberg+Nester:1980} (which we will follow).
In order to consider the initial value problem, we must consider the embedding of a family of non-overlapping, spacelike hypersurfaces $\Sigma_t$ that foliate $\mathbf{M}$, labelled by some choice of time parameter.
The geometry of this scenario has been considered extensively by Kucha\v{r}~\cite{Kuchar:1976a}; here we will merely give results that will be of some use in this work, ignoring global issues (other than the addition of surface contributions to the Hamiltonian) and assuming in all cases that a global section of the frame bundle $\mathit{GL}\mathbf{M}$ exists.

\subsection{The Surface-Adapted Basis}
\label{sect:SAB}

The spacetime metric (and inverse, respectively) may be put in surface-normal form (note the $-2$ signature of the spacetime metric)
\begin{equation}\label{eq:metric}
\mathrm{g}=\theta^\perp\otimes\theta^\perp-\gamma_{ab}\theta^a\otimes\theta^b,\quad
\mathrm{g}^{-1}=e_\perp\otimes e_\perp-\gamma^{ab}e_a\otimes e_b,
\end{equation}
where the frame and its dual coframe are given by
\begin{equation}\label{eq:basis}
e_\perp=\tfrac{1}{N}\partial_t-\tfrac{N^a}{N}e_a,\quad e_a={E^i}_a\partial_i,\quad
\theta^\perp=N dt,\quad\theta^a=N^adt+{E^a}_idx^i.
\end{equation}
The atlas fields $N$ and $N^a$ (the lapse function and shift vector respectively) play the same geometric role as that assigned to them in coordinate frame work~\cite{ADM:1959,Isenberg+Nester:1980}, namely, the shift vector is the projection of the spacetime vector field that describes the `flow of coordinate time' parallel to $\Sigma$, and the lapse function the normal projection.
(Note that they may either be thought of as a reparameterization of the spacetime metric, or as the ${E^A}_0$ components of the vielbein in a gauge where ${E^0}_i=0$ in tetrad GR.)
The only alteration of this conventional picture is that $N^a$ are the components of the shift vector in the frame $\{e_a\}$ above $\Sigma$.
It is straightforward to see that $E^{(4)}\sqrt{\lvert\mathrm{g}\rvert}=NE\sqrt{\gamma}$ where we distinguish here between the four-dimensional determinant $E^{(4)}$ and from now on write $E:=\det{{E^a}_i}$ as representing the determinant of the spatial vielbein.
We will also refer to a spatial vector as, for example $\vec{N}:=N^ae_a$.

For this surface-adapted frame we find the non-vanishing structure constants from~\eqref{eq:ST C}
\begin{subequations}\label{eq:st con}
\begin{gather}
[e_a,e_b]={C_{ab}}^ce_c,\quad
[e_\perp,e_a]={C_{\perp a}}^\perp e_\perp+{C_{\perp a}}^be_b,\quad
{C_{\perp a}}^\perp=e_a[\ln N], \\
{C_{ab}}^c={E^c}_i\bigl(e_a[{E^i}_b]-e_b[{E^i}_a]\bigr),\quad
{C_{\perp a}}^b=\frac{e_a[N^b]}{N}+\frac{N^c}{N}{C_{ac}}^b+\frac{1}{N}{E^b}_i\partial_t[{E^i}_a].
\end{gather}
\end{subequations}
Taking perpendicular and parallel projections of the compatibility conditions~\eqref{eq:COMP} results in $\Gamma^\perp_{\perp\perp}=\Gamma^\perp_{a\perp}=0$, 
$\Gamma^\perp_{\perp a}:=a_a$, $\Gamma^a_{\perp\perp}=a^a:=\gamma^{ab}a_b$,
$\Gamma^\perp_{ab}:=K_{ab}$, $\Gamma^a_{b\perp}={K^a}_{b}:=\gamma^{ac}K_{bc}$ and the surface compatibility conditions $\nabla_a[\gamma]_{bc}=0$, where $\nabla_a$ is the covariant derivative operator defined on $\Sigma$.
The remaining compatibility condition gives the relationship between the extrinsic curvature and the time derivatives of the metric and vielbein degrees of freedom
\begin{equation}\label{eq:extrinsic}
\partial_t[\gamma_{ab}]-2\gamma_{(ac}{E^c}_i\partial_t[{E^i}_{b)}] 
-2NK_{ab}-\pounds_{\! \vec{N}}[\gamma]_{ab}=0.
\end{equation}
The projections of~\eqref{eq:TORSION} ($T^\perp_{\perp a}$, $T^\perp_{ab}$, $T^a_{\perp b}$, and $T^a_{bc}$ respectively) result in $a_a=e_a[\ln N]$, $K_{[ab]}=0$, 
$\Gamma^a_{\perp b}={K^a}_{b}+{C_{\perp b}}^a$ and $\Gamma^a_{[bc]}={C_{bc}}^a$, which completes the generalization of the results of~\cite{Isenberg+Nester:1980}.

The projected components of the Ricci tensor~\eqref{eq:RICCI} that will appear in the surface reduction of~\eqref{eq:SGR} are (the operator $\mathrm{d}_\perp $ is defined in the following section)
\begin{subequations}
\begin{align}\label{eq:Ricci}
R^{(4)}_{\perp\perp}&=-\mathrm{d}_\perp [K]
+\nabla_a[a]^a+a^aa_a-{K^a}_{b}{K^b}_{a},\\
\label{eq:sp FE}
R^{(4)}_{ab}&=R_{ab}+\mathrm{d}_\perp [K]_{ab}
-\nabla_{(b}[a]_{a)}-a_aa_b+KK_{ab}-2K_{ac}{K^c}_{b},
\end{align}
\end{subequations}
where $K:={K^a}_a$ and we have written $R^{(4)}_{ab}$ to indicate the projected spacetime Ricci tensor and $R_{ab}$ the intrinsically defined spatial Ricci tensor.

\subsection{Surface-Adapted Derivatives}
\label{sect:SAD}

In order to consider the representation of diffeomorphisms in a frame where the metric is non-dynamical (or prescribed in some manner), we need to consider in some detail the representation of spatial diffeomorphisms.
The usual Lie action of an infinitesimal diffeomorphism of $\Sigma$ to itself generated by a vector field $\vec{X}$ may be written in an arbitrary frame as
\begin{subequations}
\begin{equation}\label{eq:surface Lie}
\pounds_{\! \vec{X}}[T^{ab\cdots}_{mn\cdots}]:=
\nabla_{\vec{X}}[T]^{ab\cdots}_{mn\cdots}
-\Delta_{\tilde{\nabla X}}[T]^{ab\cdots}_{mn\cdots},
\end{equation}
which is also applicable to densities: $\pounds_{\! \vec{X}}[\sqrt{\gamma}]=\nabla_a[\boldsymbol{X}]^a$.
In this form the outcome of the diffeomorphism is represented on the components of tensors as the sum of a covariant derivative $\nabla_{\vec{X}}:=X^a\nabla_a$ and a spatial frame rotation ${[\tilde{\nabla X}]^a}_{b}:=\nabla_b[X]^a$; in this case the frame is unaffected: $\pounds_{\! \vec{X}}[e_a]=0$.

This structure is not necessary, and while appropriate for considering a fixed frame, it is inappropriate for considering diffeomorphisms when constrained to an orthonormal frame since $\pounds_{\!\vec{X}}[\gamma]_{ab}=\nabla_{(a}[X]_{b)}\neq 0$.
Instead one may define a Lie action that operates on both the frame and tensor components as
\begin{equation}\label{eq:surface Lie prime}
\pounds^\prime_{\!\vec{X}}[e_a]=\Delta_{\tilde{\nabla X}}[e_a]
=-\nabla_a[X]^be_b
,\quad
\pounds^\prime_{\!\vec{X}}[T^{ab\cdots}_{mn\cdots}]
=\nabla_X[T]^{ab\cdots}_{mn\cdots},
\end{equation}
\end{subequations}
so that when the action on a tensor (not just the components) is considered, one finds the same result as~\eqref{eq:surface Lie}.
This representation \textit{is} more appropriate for considering orthonormal frames since the action of $\pounds^\prime$ on the components of the metric $\pounds^\prime_{\!\vec{X}}[\gamma]_{ab}=\nabla_{\vec{X}}[\gamma]_{ab}$ vanishes due to compatibility.
This is similar to how parallel transport in a principle bundle is transferred to associated bundles~\cite{Nakahara:1990} since one could just as easily define the covariant derivative to act as $\nabla_{\vec{X}}[T^{ab\cdots}_{mn\cdots}]=X^ae_a[T^{ab\cdots}_{mn\cdots}]$, and $\nabla_{\vec{X}}[e_a]=X^b\Gamma^c_{ba}e_c$.
This we do not pursue since the resulting action is not covariant with respect to general linear frame transformations when acting on frame or tensorial components separately.

Similarly, the definition of the derivative off of $\Sigma$ that is surface-covariant is defined to act on the components of tensors as~\cite{Isenberg+Nester:1980}
\begin{equation}\label{eq:J perp}
\mathrm{d}_\perp[T^{ab\cdots}_{mn\cdots}]:=e_\perp[T^{ab\cdots}_{mn\cdots}]
+\Delta_{\tilde{C}_\perp}[T]^{ab\cdots}_{mn\cdots},
\end{equation}
where $\tilde{C}_\perp$ is the matrix with components ${C_{\perp a}}^b$ defined in~\eqref{eq:st con}.
This operator defines the derivative normal to $\Sigma$ that `follows' the vielbein (since $\mathrm{d}_\perp[{E^i}_a]=0$ by definition), and describes the evolution of all quantities in terms of the original frame.
(Note that the definition of ${C_{\perp a}}^b$ involves time derivatives of the frame.)

As with the case of the Lie derivative, we may also consider the opposite case, namely where the normal derivative is defined so that the metric does not `evolve' and the frame does. 
Let us introduce an operator that allows a more general evolution of the vielbein
\begin{equation}
\mathrm{d}^\prime_\perp[{E^i}_a]:=e_\perp[{E^i}_a]
+\Delta_{\tilde{D}_\perp}[{E^i}_a]
=e_\perp[{E^i}_a]-{D_{\perp a}}^b{E^i}_b,
\end{equation}
where $\tilde{D}_\perp$ is determined by requiring that $\mathrm{d}^\prime_\perp[T^{ab\cdots}_{mn\cdots}e_a\otimes e_b\cdots \theta^m\otimes\theta^n\cdots]=\mathrm{d}_\perp[T^{ab\cdots}_{mn\cdots}e_a\otimes e_b\cdots \theta^m\otimes\theta^n\cdots]$ (so that the operators act on tensors in an identical manner).
We find that the action of $\mathrm{d}^\prime_\perp$ on the components of a tensor must be given by
\begin{equation}
\mathrm{d}^\prime_\perp[T^{ab\cdots}_{mn\cdots}]
=e_\perp[T^{ab\cdots}_{mn\cdots}]
+\Delta_{\tilde{C}_\perp^\prime}[T^{ab\cdots}_{mn\cdots}],\quad\text{where}\quad
{C^\prime_{\perp b}}^a:={C_{\perp b}}^a-{D_{\perp b}}^a
+{E^a}_ie_\perp[{E^i}_b].
\end{equation}
If we now require that $\mathrm{d}^\prime_\perp[\gamma]_{ab}=0$ (so that the normal derivative operator does not affect the components of the metric) then we may choose
\begin{equation}
{D_{\perp b}}^a=-\frac{1}{2N}\gamma^{ac}\partial_t[\gamma_{bc}]
+\frac{1}{N}e_b[N^a]
-\frac{1}{N}N^c{E^a}_ie_b[{E^i}_c],
\end{equation}
and therefore ${C^\prime_{\perp b}}^a=-\tfrac{1}{2}\gamma^{ac}e_\perp[\gamma_{bc}]$.
(This is far from unique since \textit{any} choice of $D_{\perp[ab]}$ will result in  $\mathrm{d}^\prime_\perp[\gamma]_{ab}=0$, however we have chosen what we consider to be a `simplest choice' which corresponds to the operator considered in Section~\ref{sect:Action} as well as~\cite{Clayton:1996b}.)
In terms of these operators, the compatibility condition~\eqref{eq:extrinsic} takes on either of the two forms
\begin{equation}\label{eq:metric evolution}
\mathrm{d}_\perp[\gamma]_{ab}=2K_{ab},\quad\text{or}\quad
{E^a}_i\mathrm{d}^\prime_\perp[{E^i}_b]=-{K^a}_{b}.
\end{equation}

Note that the total time derivative operators $\mathrm{d}_{t}=N\mathrm{d}_\perp +\pounds_{\! \vec{N}}$ and $\mathrm{d}^\prime_{t}=N\mathrm{d}^\prime_{\perp}+\pounds^\prime_{\!\vec{N}}$, correspond to the definition of the total derivative of the tensor, and are {\em not} equivalent to the time derivative of the components.
For example: $\mathrm{d}_{t}[X^ae_a]=\mathrm{d}^\prime_{t}[X^ae_a]=\partial_t[X^ae_a]\neq \partial_t[X^a]e_a$.

\subsection{The Hamiltonian}
\label{sect:Action}

From the results of the previous two sections, the GR action~\eqref{eq:SGR} is reduced to
\begin{equation}\label{eq:SGR decomposed}
S_{\textsc{gr}}
=\int dt\int_\Sigma d^3x\,\underline{\boldsymbol{N}}
\bigl(\mathrm{d}_\perp [K]+\gamma^{ab}\mathrm{d}_\perp [K]_{ab}
-2(\nabla_a[a]^a+a^aa_a)+R+K^2-{K^a}_{b}{K^b}_{a}\bigr).
\end{equation}
Using compatibility we find that $\nabla_a[a]^a+a^aa_a=(\gamma^{ab}\nabla_a\nabla_b[N])/N$ is a surface term in the action which will be dropped, and furthermore that
\begin{equation}\label{eq:12}
\begin{split}
\underline{\boldsymbol{N}}\bigl(\mathrm{d}_\perp [K]
+\gamma^{ab}\mathrm{d}_\perp [K]_{ab}\bigr)
&=2\partial_t[\underline{\boldsymbol{K}}]
-(\underline{K}_{ab}+\gamma_{ab}\underline{K})\partial_t[\boldsymbol{\gamma}^{ab}]
-2{\boldsymbol{K}^a}_{b}{E^b}_i\partial_t[{\underline{E}^i}_a]\\
&-2\underline{\boldsymbol{N}}^a\nabla_a[K]
-2E\nabla_a[{\boldsymbol{K}^a}_{b}N^b]
+2\underline{\boldsymbol{N}}^a\nabla_b{[K]^b}_{a}.
\end{split}
\end{equation}
Dropping the total time derivative and writing $S_{\textsc{gr}}=\int dt\,L_{\textsc{gr}}$, the Lagrangian is given by
\begin{multline}\label{eq:SGR decomp}
L_{\textsc{gr}}=\int_\Sigma d^3x\,\Bigl[
-\bigl(\underline{K}_{ab}+\gamma_{ab}\underline{K}\bigr)
\partial_t[\boldsymbol{\gamma}^{ab}]
-2{\boldsymbol{K}^a}_{b}{E^b}_i\partial_t[{\underline{E}^i}_a]\\
+\underline{\boldsymbol{N}}\bigl(R+K^2-{K^a}_{b}{K^b}_{a}\bigr)
+2\underline{\boldsymbol{N}}^a\bigl(\nabla_b{[K]^b}_{a}-\nabla_a[K]\bigr)\Bigr].
\end{multline}

In this form the Lagrangian is most easily treated in Palatini form, that is, by considering the extrinsic curvature $K_{ab}$ as a tensor of Lagrange multipliers that enforce~\eqref{eq:extrinsic}.
The configuration of the system at any instant in time is described by a section of $\mathit{Riem}_\Sigma\times \mathit{GL}\Sigma$, however due to the form of~\eqref{eq:SGR decomp} it is convenient to choose instead the densitized canonical coordinates $\boldsymbol{\gamma}^{ab}$ and ${\underline{E}^i}_a$, and so we are considering sections of $\boldsymbol{\mathit{Riem}}_\Sigma\times\underline{\mathit{GL}\Sigma}$.
Determining the conjugate momenta via
\begin{equation}\label{eq:Conjugate}
\underline{\pi}_{ab}:=
\frac{\delta L_{\textsc{gr}}}{\partial_t[\boldsymbol{\gamma}^{ab}]}
=-(\underline{K}_{ab}+\gamma_{ab}\underline{K}),\quad
{\boldsymbol{p}^a}_i:=\frac{\delta L_{\textsc{gr}}}{\partial_t[{\underline{E}^i}_a]}
=-2{\boldsymbol{K}^a}_{b}{E^b}_i,
\end{equation}
the set of canonical coordinate--momentum pairs is
\begin{equation}\label{eq:CCs}
\bigl\{(Q^I,P_I)\bigr\}:=
\bigl\{(\boldsymbol{\gamma}^{ab},\underline{\pi}_{ab}),
({\underline{E}^i}_a,{\boldsymbol{p}^a}_i)\bigr\},
\end{equation}
corresponding to the phase space $T^*\bigl(\boldsymbol{\mathit{Riem}}_\Sigma\times\underline{\mathit{GL}\Sigma})$.
From the canonical symplectic form on this phase space comes the standard form of the Poisson bracket
\begin{equation}\label{eq:standard Pb}
\{F,G\}:=\int_\Sigma d^3x\,\sum_I
\biggl(\frac{\delta F}{\delta Q^I(x)}\frac{\delta G}{\delta P_I(x)}
-\frac{\delta F}{\delta P_I(x)}\frac{\delta G}{\delta Q^I(x)}\biggr),
\end{equation}
where $F$ and $G$ are functions of the phase space variables.
(The results herein could be globalized along the lines of~\cite{Fischer+Marsden:1979} in which case we would introduce the related weak symplectic form on phase space, however a local treatment is sufficient for this work.)
In addition to the phase space coordinates we have the Lagrange multipliers $\bigl(N,N^a,{N^a}_{b},K_{ab},\lambda^{ab}\bigr)$ and the Hamiltonian determined from $H_{\textsc{gr}}=\int_\Sigma dx\,\sum_I\dot{Q}^IP_I-L_{\textsc{gr}}$ is written as
\begin{equation}\label{eq:H star1}
\begin{split}
\mathcal{H}_{\textsc{gr}}
&=\underline{\boldsymbol{\lambda}}^{ab}
\bigl(K_{ab}-\bar{a}k_{ab}-(1-\bar{a})k^\prime_{ab}\bigr)
+{N^b}_{a}{\underline{\boldsymbol{\mathcal{J}}}^a}_{b}\\
&+\underline{\boldsymbol{N}}\bigl(-R+{K^a}_{b}{K^b}_{a}-K^2\bigr)
+2\underline{\boldsymbol{N}}^a\nabla_b\bigl[-{K^b}_{a}+\delta^b_a K\bigr],
\end{split}
\end{equation}
where we have defined 
\begin{equation}
k_{ab}:=-(\pi_{ab}-\tfrac{1}{4}\gamma_{ab}\pi),\quad
k^\prime_{ab}:=-\tfrac{1}{2}\gamma_{(ac}{p^c}_i{E^i}_{b)}.
\end{equation}

The tensor of constraints enforced by ${N^a}_{b}$ (defining the scalars $\pi:=\gamma^{ab}\pi_{ab}$ and $p:={p^a}_i{E^i}_a$):
\begin{equation}\label{eq:GL3R constraints}
{\underline{\boldsymbol{\mathcal{J}}}^a}_{b}:=
2\boldsymbol{\gamma}^{ac}\underline{\pi}_{bc}
-\underline{\boldsymbol{\pi}}\delta^a_b
-{\boldsymbol{p}^a}_i{\underline{E}^i}_b
+\underline{\boldsymbol{p}}\delta^a_b,
\end{equation}
guarantee the consistency of~\eqref{eq:Conjugate}, and are uniquely specified by the requirement that they generate infinitesimal $\mathfrak{gl}(3,\mathbb{R})$ frame rotations on phase space.
It is straightforward to show that 
\begin{subequations}
\begin{gather}\label{eq:PB-GL3R}
\bigl\{\boldsymbol{\gamma}^{ab},
{\underline{\boldsymbol{\mathcal{J}}}^c}_{d}[{\omega^d}_{c}]\bigr\}=
\Delta_{\tilde{\omega}}[\boldsymbol{\gamma}]^{ab},\quad
\bigl\{\underline{\pi}_{ab},
{\underline{\boldsymbol{\mathcal{J}}}^c}_{d}[{\omega^d}_{c}]
\bigr\}=
\Delta_{\tilde{\omega}}[\underline{\pi}]_{ab}, \\
\bigl\{{\underline{E}^i}_a,
{\underline{\boldsymbol{\mathcal{J}}}^b}_{c}[{\omega^c}_{b}]
\bigr\}=
\Delta_{\tilde{\omega}}[{\underline{E}^i}_a],\quad
\bigl\{{\boldsymbol{p}^a}_i,
{\underline{\boldsymbol{\mathcal{J}}}^b}_{c}[{\omega^c}_{b}]
\bigr\}=
\Delta_{\tilde{\omega}}[{\boldsymbol{p}^a}_i], 
\end{gather}
\end{subequations}
where the infinitesimal $\mathrm{GL}(3,\mathbb{R})$ frame rotations $\Delta_{\tilde{\omega}}$ are defined in~\eqref{eq:infinit}, and we have made use of the notation, for example ${\underline{\boldsymbol{\mathcal{J}}}^b}_{a}[{\omega^a}_{b}]:=\int_\Sigma d^3x\; {\omega^a}_{b}{\underline{\boldsymbol{\mathcal{J}}}^b}_{a}$.
These generators (the phase space representation of $\Delta^a_b$ of~\cite{Clayton:1996b}) satisfy the $\mathfrak{gl}(3,\mathbb{R})$ Lie algebra
\begin{equation}\label{eq:gL3r Lie}
\bigl\{{\underline{\boldsymbol{\mathcal{J}}}^a}_{b},
{\underline{\boldsymbol{\mathcal{J}}}^c}_{d}[{\omega^d}_{c}]\bigr\}=
\Delta_{\tilde{\omega}}{\bigl[\underline{\boldsymbol{\mathcal{J}}}\bigr]^a}_{b},
\end{equation}
strongly on phase space.
This differs from~\eqref{eq:lie left} by a sign due to the fact that the operator $\Delta$ acts from the left while $\{\cdot,\mathcal{J}\}$ acts from the right.
It is important to note that $\mathcal{J}^{ab}$ is \textit{not} a symmetric tensor of constraints, in fact it is precisely the antisymmetric components that generate $\mathrm{SO}(3)$ rotations in a local orthonormal frame.

The constraints imposed by $\lambda^{ab}$ explicitly relate the extrinsic curvature to the conjugate momenta through
\begin{equation}\label{eq:ks}
{K}_{ab}\approx
\bar{a}k_{ab}
+(1-\bar{a})k^\prime_{ab}, 
\end{equation}
and in the form given, the variation of~\eqref{eq:H star1} with respect to $K_{ab}$ would then determine $\lambda^{ab}$ in terms of the other Lagrange multipliers as $\lambda_{ab}\approx -2N(K_{ab}-\gamma_{ab}K)-2(\nabla_{(a}[N]_{b)}-\gamma_{ab}\nabla_c[N]^c)$.
It is rather inconvenient to carry around the Lagrange multipliers $K_{ab}$ and $\lambda^{ab}$ solely to deal with the ambiguity in determining $K_{ab}$ from the conjugate momenta.
This may be circumvented in a simple manner.

When passing to the Hamiltonian $H_{\textsc{gr}}$ we may replace each occurrence of the extrinsic curvature in the Hamiltonian with some combination of the canonical momenta consistent with~\eqref{eq:ks}, with the result that the only place $K_{ab}$ will occur is in the constraint enforced by $\lambda^{ab}$.
The variation of $K_{ab}$ will then enforce $\lambda^{ab}\approx 0$, and neither $K_{ab}$ nor $\lambda^{ab}$ will play any further role in the Hamiltonian system.
Therefore this term may be consistently dropped from the Hamiltonian and these Lagrange multipliers are removed.
The Hamiltonian determined in this way is identical to that which would occur if one had replaced the extrinsic curvature by~\eqref{eq:extrinsic} in the Lagrangian~\eqref{eq:SGR decomp}.
In this way we see that the GR Hamiltonian can always be written as
\begin{subequations}\label{eq:general forms}
\begin{equation}\label{eq:general form}
H_{\textsc{gr}}=\int_\Sigma d^3x\bigl(
N\underline{\boldsymbol{\mathcal{H}}}
+N^a\underline{\boldsymbol{\mathcal{H}}}_a
+{N^a}_{b}{\underline{\boldsymbol{\mathcal{J}}}^b}_{a}
\bigr)+E_{\textsc{gr}},
\end{equation}
where in the Hamiltonian and momentum constraints
\begin{equation}\label{eq:generic constraints}
\underline{\boldsymbol{\mathcal{H}}}:=
-\underline{\boldsymbol{R}}
+\boldsymbol{E}({K^a}_b{K^b}_a-K^2),\quad
\underline{\boldsymbol{\mathcal{H}}}_a:=
-2\boldsymbol{E}\nabla_b[{K^b}_a-\delta^b_a K],
\end{equation}
\end{subequations}
$K_{ab}$ represents some linear combination of $k_{ab}$ and $k^\prime_{ab}$ consistent with~\eqref{eq:ks}.

In order for these field equations to properly follow from the Hamiltonian, it was necessary to add the surface contributions $E_{\text{\textsc{gr}}}$ to the Hamiltonian~\eqref{eq:general form}, which are required to satisfy~\cite{Regge+Teitelboim:1974}
\begin{subequations}\label{eq:surface terms}
\begin{equation}\label{eq:delta surface}
\delta E_{\text{\textsc{gr}}}+S_R+S_k=0
\end{equation}
in order for the field equations to be properly recovered from the Hamiltonian, where
\begin{equation}
S_R:=\int_{\partial\Sigma} d\underline{S}_a\,
\bigl(N\nabla_b[\delta\boldsymbol{S}^{ab}]
-\nabla_b[N]\delta\boldsymbol{S}^{ba}\bigr),
\end{equation}
comes from the variations of the surface Ricci curvature term, and ($H_{ab}:=-2(K_{ab}-\gamma_{ab}K)$, and $K_{ab}$ is again determined in terms of $k_{ab}$ and $k^\prime_{ab}$)
\begin{equation}\label{eq:k surface}
\begin{split}
S_k:&=-\tfrac{1}{2}\int_{\partial\Sigma} d\underline{S}_a
N^a\boldsymbol{H}^{bc}\delta\gamma_{bc}
+\int_{\partial\Sigma} d\underline{S}_a N^b\delta{\boldsymbol{H}^a}_{b}\\
&+\int_{\partial\Sigma} d\underline{S}_a\bigl[
N^a{\boldsymbol{H}^c}_{b}{E^b}_i
-N^b{\boldsymbol{H}^a}_{b}{E^c}_i
+N^b{\boldsymbol{H}^c}_{b}{E^a}_i
+N_b\boldsymbol{H}^{ac}{E^b}_i\bigr]\delta{E^i}_c,
\end{split}
\end{equation}
\end{subequations}
comes from variations of the momentum constraint.

\subsection{The Constraint Algebras}
\label{sect:constraints}

In this section we will consider two choices of the diffeomorphism constraints.
First we will choose $\underline{\boldsymbol{\mathcal{H}}}$ and $\underline{\boldsymbol{\mathcal{H}}}_a$ to act on phase space as the operators $\pounds$ and $\mathrm{d}_\perp$ from Section~\ref{sect:SAD}, and refer to the Hamiltonian constructed from these as $H_{\textsc{gr}}$.
This corresponds to $\bar{a}=1$ in~\eqref{eq:ks}, and $K_{ab}$ is replaced everywhere by $k_{ab}$ in~\eqref{eq:general forms} and~\eqref{eq:surface terms}.
This choice will closely resemble the coordinate approach to canonical GR and has a straightforward coordinate frame gauge reduction.
Since these constraints have a nontrivial action on the spatial metric, it is nontrivial to recover the limit where only orthonormal spatial vielbeins are considered. 
This is instead accomplished by considering a second set of generators $\underline{\boldsymbol{\mathcal{H}}}^\prime$ and $\underline{\boldsymbol{\mathcal{H}}}^\prime_a$ which act on phase space as $\pounds^\prime$ and $\mathrm{d}^\prime_\perp$ from Section~\ref{sect:SAD}, and a second Hamiltonian $H^\prime_{\textsc{gr}}$ is defined by attaching primes to the constraints in~\eqref{eq:general forms}.
In this scheme $K_{ab}$ is replaced everywhere by $k^\prime_{ab}$ (corresponding to $\bar{a}=0$), and there is also mixing between the $\mathrm{GL}(3,\mathbb{R})$ constraints and the Momentum constraint appearing in~\eqref{eq:general forms}.
This second system will have a straightforward reduction to canonical vierbein GR, however the specialization to spatial coordinate frames is nontrivial.

The Hamiltonian $H_{\textsc{gr}}$ is constructed from~\eqref{eq:GL3R constraints} and the operators that act on phase space as $\mathrm{d}_\perp$ and $\pounds$ respectively:
\begin{subequations}\label{eq:unprimed constraints}
\begin{align}
\underline{\boldsymbol{\mathcal{H}}}:&=
-\underline{\boldsymbol{R}}
+\boldsymbol{E}\bigl({k^a}_b{k^b}_a-k^2\bigr),\\
\label{eq:mom unprimed}
\underline{\boldsymbol{\mathcal{H}}}_a:&=
\boldsymbol{E}\nabla_b[2\gamma^{bc}\pi_{ac}-\delta^b_a\pi]
=-2\boldsymbol{E}\nabla_b[{k^b}_a-\delta^b_a k].
\end{align}
\end{subequations}
From these we compute 
\begin{subequations}\label{eq:surface diffeos}
\begin{gather}
\bigl\{\boldsymbol{\gamma}^{ab},
\underline{\boldsymbol{\mathcal{H}}}_c[f^c]\bigr\}
=\pounds_{\! \vec{f}}[\boldsymbol{\gamma}]^{ab},\quad
\bigl\{\underline{\pi}_{ab},
\underline{\boldsymbol{\mathcal{H}}}_c[f^c]\bigr\}
=
E\pounds_{\! \vec{f}}[\pi]_{ab},\quad
\label{eq:diff frame}
\bigl\{{\underline{E}^i}_a,
\underline{\boldsymbol{\mathcal{H}}}_c[f^c]\bigr\}
=0,\\
\label{eq:diff p}
\bigl\{{\boldsymbol{p}^a}_i,
\underline{\boldsymbol{\mathcal{H}}}_c[f^c]\bigr\}
=-{E^b}_i\bigl(f^a\underline{\boldsymbol{\mathcal{H}}}_b
-\tfrac{1}{2}\delta^a_b f^c\underline{\boldsymbol{\mathcal{H}}}_c\bigr)
+{E^b}_i\pounds_{\! \vec{f}}[
2\boldsymbol{\gamma}^{ac}\pi_{bc}-\tfrac{1}{2}\delta^a_b\boldsymbol{\pi}]
\approx
{E^b}_i\pounds_{\! \vec{f}}[{\boldsymbol{p}^a}_j{E^j}_b],
\end{gather}
\end{subequations}
and (defining $\nabla^2:=\gamma^{ab}\nabla_a\nabla_b$)
\begin{subequations}\label{eq:H perp}
\begin{align}
\label{eq:H gamma}
\bigl\{\boldsymbol{\gamma}^{ab},\underline{\boldsymbol{\mathcal{H}}}[f]\bigr\}
&=-\boldsymbol{f}\bigl(2k^{ab}-\gamma^{ab}k\bigr),\quad
\bigl\{{\underline{E}^i}_a,\underline{\boldsymbol{\mathcal{H}}}[f]\bigr\}
=0,\\
\label{eq:H pi}
\bigl\{\underline{\pi}_{ab},\underline{\boldsymbol{\mathcal{H}}}[f]\bigr\}
&=-E\bigl(\nabla_{a}\nabla_{b}[f]
+\gamma_{ab}\nabla^2[f]\bigr)
+f\underline{R}_{ab}\nonumber \\
&-\underline{f}
\bigl(2k_{ac}{k^c}_b+kk_{ab}-\gamma_{ab}k^2\bigr)
+\underline{f}\gamma_{ab}
\bigl({k^c}_d{k^d}_c-k^2\bigr),\\
\label{eq:H p}
\bigl\{{\boldsymbol{p}^a}_i,\underline{\boldsymbol{\mathcal{H}}}[f]\bigr\}
&=-2{E^b}_i\boldsymbol{\gamma}^{ac}\nabla_{c}\nabla_{b}[f]
+f{E^b}_i\boldsymbol{\gamma}^{ac}
\bigl(2R_{cb}
-\tfrac{1}{2}\gamma_{cb}R\bigr)\nonumber \\
&+\tfrac{1}{2}{E^a}_i\boldsymbol{f}\bigl({k^c}_d{k^d}_c-k^2\bigr).
\end{align}
\end{subequations} 
It is straightforward to check from these that $\bigl\{k_{ab},\underline{\boldsymbol{\mathcal{H}}}[N]\bigr\}
\approx -NR_{ab}+2Nk_{ac}{k^c}_{b}-Nkk_{ab}+\nabla_{a}\nabla_{b}[N]$, and we therefore reproduce the field equations~\eqref{eq:sp FE}.
We also find (using~\eqref{eq:H gamma}) that $\bigl\{\gamma_{ab},\underline{\boldsymbol{\mathcal{H}}}[N]\bigr\}=2Nk_{ab}$, reproducing the dynamical compatibility condition~\eqref{eq:extrinsic} or, equivalently,~\eqref{eq:metric evolution}.

Using the above results we compute the constraint algebra
\begin{subequations}\label{eq:unprimed algebra}
\begin{align}
\bigl\{\underline{\boldsymbol{\mathcal{H}}},
{\underline{\boldsymbol{\mathcal{J}}}^a}_{b}[{\omega^b}_{a}]\bigr\}
&=0,\quad
\bigl\{\underline{\boldsymbol{\mathcal{H}}}_a,
{\underline{\boldsymbol{\mathcal{J}}}^b}_{c}[{\omega^c}_{b}]\bigr\}=
\Delta_{\tilde{\omega}}[\underline{\boldsymbol{\mathcal{H}}}]_a,\\
\bigl\{\underline{\boldsymbol{\mathcal{H}}}[f],
\underline{\boldsymbol{\mathcal{H}}}[g]\bigr\}
&=\int_{\Sigma} d^3x\,\gamma^{ab}\bigl(f\nabla_a[g]-g\nabla_a[f]\bigr)
\underline{\boldsymbol{\mathcal{H}}}_b,\\
\bigl\{\underline{\boldsymbol{\mathcal{H}}}[f],
\underline{\boldsymbol{\mathcal{H}}}_c[g^c]\bigr\}
&=\int_{\Sigma} d^3x\,\underline{f}\pounds_{\! \vec{g}}[\boldsymbol{\mathcal{H}}],\\
\bigl\{\underline{\boldsymbol{\mathcal{H}}}_a[f^a],
\underline{\boldsymbol{\mathcal{H}}}_a[g^a]\bigr\}
&=\int_{\Sigma} d^3x\,\underline{f}^a\pounds_{\! \vec{g}}[\boldsymbol{\mathcal{H}}]_a,
\end{align}
\end{subequations}
which, in the limit of a spatial coordinate frame (${E^i}_a=\delta^i_a$), agrees with the Dirac algebra~\cite{Teitelboim:1973,Teitelboim:1980,Hojman+Kuchar+Teitelboim:1976,Fischer+Marsden:1979}.

The Hamiltonian $H^\prime_{\textsc{gr}}$ is constructed from~\eqref{eq:GL3R constraints} and the operators that act on phase space as $\mathrm{d}^\prime_\perp$ and $\pounds^\prime$ respectively:
\begin{subequations}\label{eq:primed constraints}
\begin{align}
\underline{\boldsymbol{\mathcal{H}}}^\prime:
&=-\underline{\boldsymbol{R}}
+\boldsymbol{E}
\bigl({{k^\prime}^a}_b{{k^\prime}^b}_a-{k^\prime}^2\bigr)
=\underline{\boldsymbol{\mathcal{H}}}
+k^\prime_{ab}\underline{\boldsymbol{\mathcal{J}}}^{ab}
-\tfrac{1}{4}\bigl(\mathcal{J}^{(ab)}\underline{\boldsymbol{\mathcal{J}}}_{(ab)}
-\tfrac{1}{2}{\mathcal{J}^a}_{a}
{\underline{\boldsymbol{\mathcal{J}}}^b}_{b}\bigr),\\
\label{eq:mom primed}
\underline{\boldsymbol{\mathcal{H}}}^\prime_a:&=
\boldsymbol{E}\nabla_b[{p^b}_i{E^i}_a-\delta^b_a p]
=-2\boldsymbol{E}\nabla_b[{{k^\prime}^b}_a-\delta^b_ak^\prime]
+E\gamma_{ab}\nabla_c[\boldsymbol{\mathcal{J}}]^{[bc]}
=\underline{\boldsymbol{\mathcal{H}}}_a
-E\nabla_b{[\boldsymbol{\mathcal{J}}]^b}_{a},
\end{align}
\end{subequations}
where we have indicated the relationship between these primed constraints and the unprimed constraints in~\eqref{eq:unprimed constraints}.
Note that in~\eqref{eq:mom primed} there is a nontrivial mixing of the $\mathrm{GL}(3,\mathbb{R})$ constraints and the momentum constraint as it appears in~\eqref{eq:general forms}.
These act as
\begin{subequations}\label{eq:surface diffeos II}
\begin{gather}
\label{eq:diff frame II}
\bigl\{\boldsymbol{\gamma}^{ab},
\underline{\boldsymbol{\mathcal{H}}}^\prime_c[f^c]\bigr\}
=0,\quad
\bigl\{{\underline{E}^i}_a,
\underline{\boldsymbol{\mathcal{H}}}^\prime_c[f^c]\bigr\}
=\Delta_{\tilde{\nabla{f}}}[{\underline{E}^i}_a],\\
\bigl\{\underline{\pi}_{ab},
\underline{\boldsymbol{\mathcal{H}}}^\prime_c[f^c]\bigr\}
=E\bigl(V_{ab}[f]-\gamma_{ab}{V^c}_{c}[f]\bigr)
\approx \pounds^\prime_{\!\vec{f}}[\underline{\pi}]_{ab},\\
\label{eq:diff p II}
\bigl\{{\boldsymbol{p}^a}_i,
\underline{\boldsymbol{\mathcal{H}}}^\prime_c[f^c]\bigr\}
=\Delta_{\tilde{\nabla f}}[{\boldsymbol{p}^a}_i]
+{E^b}_i{\boldsymbol{U}^a}_{b}[\vec{f}]
-\tfrac{1}{2}{E^a}_i{\boldsymbol{U}^c}_{c}[\vec{f}]
\approx\pounds^\prime_{\!\vec{f}}[{\boldsymbol{p}^a}_i],
\end{gather}
\end{subequations}
where (${H^a}_{b}:={p^a}_i{E^i}_b-\delta^a_b p$)
\begin{subequations}
\begin{align}
V_{ab}[f]:&=\tfrac{1}{2}\nabla^c[f_aH_{cb}
+f_cH_{ab}-f_aH_{bc}],\\
{\boldsymbol{U}^a}_{b}[\vec{f}]:&=
\nabla_b[f]^c{\boldsymbol{H}^a}_{c}
+\gamma_{bd}\nabla_c\bigl[f^c\boldsymbol{H}^{(ad)}\bigr]
-\gamma_{bd}\nabla_c\bigl[f^a\boldsymbol{H}^{(cd)}\bigr]
+\nabla_c\bigl[f_b\boldsymbol{H}^{[ca]}\bigr],
\end{align}
\end{subequations}
and
\begin{subequations}\label{eq:H perp prime}
\begin{align}
\label{eq:H gamma II}
\bigl\{\boldsymbol{\gamma}^{ab},\underline{\boldsymbol{\mathcal{H}}}^\prime[f]\bigr\}
&=0,\quad
\bigl\{{\underline{E}^i}_a,\underline{\boldsymbol{\mathcal{H}}}^\prime[f]\bigr\}
=f\Delta_{\tilde{k}^\prime}[{\underline{E}^i}_a],\\
\label{eq:H pi II}
\bigl\{\underline{\pi}_{ab},\underline{\boldsymbol{\mathcal{H}}}^\prime[f]\bigr\}
&=-E\bigl(\nabla_{a}\nabla_{b}[f]
+\gamma_{ab}\nabla^2[f]\bigr)
+f\underline{R}_{ab}
+\underline{f}\gamma_{ab}
\bigl({{k^\prime}^c}_d{{k^\prime}^d}_c-{k^\prime}^2\bigr),\\
\label{eq:H p II}
\bigl\{{\boldsymbol{p}^a}_i,\underline{\boldsymbol{\mathcal{H}}}^\prime[f]\bigr\}
&=-2{E^b}_i\boldsymbol{\gamma}^{ac}\nabla_{c}\nabla_{b}[f]
+f{E^b}_i\boldsymbol{\gamma}^{ac}
\bigl(2R_{cb}
-\tfrac{1}{2}\gamma_{cb}R\bigr)\nonumber \\
&+\boldsymbol{f}\bigl({{k^\prime}^a}_b-\delta^a_bk^\prime\bigr){p^b}_i
+\tfrac{1}{2}\boldsymbol{f}{E^a}_i
\bigl({{k^\prime}^c}_d{{k^\prime}^d}_c-{k^\prime}^2\bigr).
\end{align}
\end{subequations} 
Here we find that~\eqref{eq:diff frame II} reproduces~\eqref{eq:metric evolution} and that~\eqref{eq:sp FE} is satisfied when written in terms of the operator $\mathrm{d}^\prime_\perp$.

Using these results we compute the primed constraint algebra:
\begin{subequations}\label{eq:primed algebra}
\begin{align}
\bigl\{\underline{\boldsymbol{\mathcal{H}}}^\prime,
{\underline{\boldsymbol{\mathcal{J}}}^a}_{b}[{\omega^b}_{a}]\bigr\}
&=0,\quad
\bigl\{\underline{\boldsymbol{\mathcal{H}}}^\prime_a,
{\underline{\boldsymbol{\mathcal{J}}}^b}_{c}[{\omega^c}_{b}]\bigr\}=
\Delta_{\tilde{\omega}}[\underline{\boldsymbol{\mathcal{H}}}^\prime]_a,\\
\bigl\{\underline{\boldsymbol{\mathcal{H}}}^\prime[f],
\underline{\boldsymbol{\mathcal{H}}}^\prime[g]\bigr\}
&=\int_{\Sigma} d^3x\,\bigl(f\nabla_a[g]-g\nabla_a[f]\bigr)
\Bigl(\gamma^{ab}\underline{\boldsymbol{\mathcal{H}}}^\prime_b
-E\nabla_b[\boldsymbol{\mathcal{J}}]^{[ab]}\Bigr)
,\\
\label{eq:mixed II}
\bigl\{\underline{\boldsymbol{\mathcal{H}}}^\prime[f],
\underline{\boldsymbol{\mathcal{H}}}^\prime_c[g^c]\bigr\}
&=\int_{\Sigma} d^3x\,\Bigl(
\underline{f}\pounds_{\!\vec{g}}[\boldsymbol{\mathcal{H}}^\prime]
-fg^a\Delta_{\tilde{k}^\prime}
[\underline{\boldsymbol{\mathcal{H}}}^\prime]_a
+2g^a\nabla_c\bigl[fk^\prime_{ab}\bigr]
\underline{\boldsymbol{\mathcal{J}}}^{[bc]}\Bigr),\\
\bigl\{\underline{\boldsymbol{\mathcal{H}}}^\prime_a[f^a],
\underline{\boldsymbol{\mathcal{H}}}^\prime_a[g^a]\bigr\}
&=-\int_{\Sigma} d^3x\,f^ag^b{R^c}_{dab}
{\underline{\boldsymbol{\mathcal{J}}}^d}_{c},
\end{align}
\end{subequations}
which, when specialized to an orthonormal spatial frame ($\gamma_{ab}=\delta_{ab}$), is equivalent to that in~\cite{Henneaux:1983,Charap+Henneaux+Nelson:1988}.

These strong results agree with~\cite{Clayton:1996b} up to the overall sign attributable to the fact that we are considering a right (Poisson) action in this work, whereas in~\cite{Clayton:1996b} we derived the left action~\cite{Hojman+Kuchar+Teitelboim:1976}.
(Due to an unfortunate typesetting error, the left hand sides of equations~($28$b-e) of~\cite{Clayton:1996b} should read $[\Delta_{ax},\Delta_y]$, $[\delta_{\mathsf{n}x},\Delta_y]$, $[\delta_{\mathsf{n}x},\Delta_{ay}]$, and $[\delta_{ax},\Delta_y]$ respectively.
Also note the rather obvious typo in the last line of~($10$), which should read $\nabla_b[k]^a_c-\mathrm{g}^{ad}\nabla_d[k]_{bc}$.)
Either of the algebras~\eqref{eq:unprimed algebra} or~\eqref{eq:primed algebra} is a local representation of the Lie algebra $\mathrm{L}\mathit{Diff}_0\mathbf{M}$ (the connected component since the exponential map is not onto) of the spacetime diffeomorphism group $\mathit{Diff}\mathbf{M}$.
Note that although they were derived from a $(3+1)$ action, that algebras are valid in any dimension.

The Hamiltonian $H_{\text{\textsc{gr}}}$ generates the weak evolution equations
\begin{subequations}\label{eq:dots}
\begin{align}
\label{eq:gamma dot}
\partial_t[\boldsymbol{\gamma}^{ab}]=\{\boldsymbol{\gamma}^{ab},H_{\textsc{gr}}\}
\approx& \Delta_{\tilde{N}}[\boldsymbol{\gamma}]^{ab}
+\pounds_{\! \vec{N}}[\boldsymbol{\gamma}]^{ab}
-N\Delta_{\tilde{k}}[\boldsymbol{\gamma}]^{ab},\\
\partial_t[{\underline{E}^i}_a]=\{{\underline{E}^i}_a,H_{\textsc{gr}}\}
\approx& \Delta_{\tilde{N}}[{\underline{E}^i}_a],\\
\partial_t[\underline{\pi}_{ab}]
=\{\underline{\pi}_{ab},H_{\textsc{gr}}\}
\approx& \Delta_{\tilde{N}}[\underline{\pi}]_{ab}
+E\pounds_{\! \vec{N}}[\pi]_{ab}
-E\bigl(\nabla_{a}\nabla_{b}[N]
+\gamma_{ab}\nabla^2[N]\bigr)\nonumber \\
&+N\underline{R}_{ab}
-NE(2k_{ac}{k^c}_{b}+kk_{ab})
+NE\gamma_{ab}{k^c}_{d}{k^d}_{c},\\
\partial_t[{\boldsymbol{p}^a}_i]=\{{\boldsymbol{p}^a}_i,H_{\textsc{gr}}\}
\approx& \Delta_{\tilde{N}}[{\boldsymbol{p}^a}_i]
-2{E^b}_i\pounds_{\! \vec{N}}[{\boldsymbol{k}^a}_{b}]
-2{E^b}_i\boldsymbol{\gamma}^{ac}\nabla_{b}\nabla_{c}[N]
+2N{E^b}_i\boldsymbol{\gamma}^{ac}R_{cb}.
\end{align}
\end{subequations}
Using metric compatibility,~\eqref{eq:gamma dot} may be written as
\begin{equation}\label{eq:mult choice}
\Delta_{\tilde{\omega}}[\boldsymbol{\gamma}]^{ab},\quad\text{where}\quad
{\omega^a}_{b}={N^a}_{b}-\nabla_b[N]^a-N{k^a}_{b},
\end{equation}
which is consistent with the easily verified result (noting that $\mathcal{J}_{(ab)}\approx 0$ is equivalent to $k_{ab}\approx k^\prime_{ab}$) that the weak evolution equations for the primed system with Hamiltonian written as 
$H^\prime_{\text{\textsc{gr}}}=\int_\Sigma d^3x\; 
\bigl(N^\prime\underline{\boldsymbol{\mathcal{H}}}^\prime
+{N^\prime}^a\underline{\boldsymbol{\mathcal{H}}}^\prime_a
+{{N^\prime}^a}_b{{\underline{\boldsymbol{\mathcal{J}}}}^b}_a\bigr)$ is equivalent to inserting
\begin{equation}
N=N^\prime,\quad
N^a={N^\prime}^a,\quad
{N^a}_{b}={{N^\prime}^a}_{b}+\nabla_b[N^\prime]^a+N^\prime{k^a}_{b},
\end{equation}
in~\eqref{eq:dots}.
In either of these formulations, the system is parameterized by the $30$ fields~\eqref{eq:CCs}, with $13$ primary, first--class constraints (${\underline{\boldsymbol{\mathcal{J}}}^a}_{b}$ with either $\underline{\boldsymbol{\mathcal{H}}}$ and $\underline{\boldsymbol{\mathcal{H}}}_a$, or $\underline{\boldsymbol{\mathcal{H}}}^\prime$ and $\underline{\boldsymbol{\mathcal{H}}}_a^\prime$), and the $13$ associated Atlas Lagrange multiplier fields ($N$, $N^a$ and ${N^a}_{b}$), so that we find two configuration space degrees of freedom per spacetime point as expected~\cite{Isenberg+Nester:1980}.

Beginning from the weak evolution equations~\eqref{eq:dots}, choosing ${N^a}_{b}\approx 0$ results in a system in which the spatial vielbeins do not evolve, and solving the $\mathrm{GL}(3,\mathbb{R})$ constraints by choosing ${p^a}_i\approx{E^b}_i(2\gamma^{ac}\pi_{bc}-\tfrac{1}{2}\delta^a_b\pi)$ we can consider the evolution of $\bigl(\boldsymbol{\gamma}^{ab},\underline{\pi}_{ab}\bigr)$ sector exclusively.
Instead choosing ${{N^\prime}^a}_{b}=0$, we find a system in which $\boldsymbol{\gamma}^{ab}$ does not evolve and one may play the opposite game, solving the $\mathrm{GL}(3,\mathbb{R})$ constraints by $\pi_{ab}\approx \tfrac{1}{2}\gamma_{ac}({p^c}_i{E^i}_b+\delta^c_bp)$ and consider the evolution of $\bigl({\boldsymbol{p}^a}_i,{\underline{E}^i}_a\bigr)$.
Note that from~\eqref{eq:mult choice} one need only require that 
\begin{equation}\label{eq:choice 2}
N^{(ab)}=\gamma^{c(a}\nabla_c[N]^{b)}
-N\bigl(\pi^{ab}-\tfrac{1}{4}\gamma^{ab}\pi\bigr)
=\gamma^{c(a}\nabla_c[N]^{b)}+Nk^{ab},
\end{equation}
in order to find a system in which the metric degrees of freedom do not evolve, allowing other gauge choices in this case.

One may treat \textit{any} choice of initial data in either of these two ways.
By transforming the initial data by $\Delta_{\tilde{\omega}}$ defined so that $\Delta_{\tilde{\omega}}[{E^i}_a]=\delta^i_a$ (which is uniquely determined from the inverse as ${[\tilde{\omega}]^a}_{i}={E^a}_i$), we transform the initial data to a physically equivalent set for which the frame is holonomic.
If instead we make a choice of $\omega$ such that $\Delta_{\tilde{\omega}}[\gamma]^{ab}=\delta^{ab}$ ($\tilde{\omega}$ is defined only up to arbitrary spatial rotations which are generated by $\mathcal{J}^{[ab]}$) we may diagonalize the spatial metric.

\section{Conclusions}
\label{sect:concl}

We have shown that by treating the components of the metric and the frame on an equal footing we not only have a consistent variational principle, but the generalized setting encompasses both the coordinate and orthonormal frame approaches via a choice of gauge.
From the Einstein--Hilbert action for general relativity we have derived a Hamiltonian in the (extended) standard form $H_{\textsc{gr}}=\int_\Sigma dx\;\bigl(N\underline{\boldsymbol{\mathcal{H}}}+N_a\underline{\boldsymbol{\mathcal{H}}}_a+{N^a}_{b}{\underline{\boldsymbol{\mathcal{J}}}^b}_{a}\bigr)$, where $\underline{\boldsymbol{\mathcal{H}}}$ and $\underline{\boldsymbol{\mathcal{H}}}_a$ are the generators (of the connected component) of spacetime diffeomorphisms, and ${\underline{\boldsymbol{\mathcal{J}}}^a}_{b}$ are the generators of $\mathfrak{gl}(3,\mathbb{R})$ which generate infinitesimal changes of frame on the spatial hypersurface $\Sigma$.
The resulting constraint algebra was computed and found to be in agreement with the algebra derived by purely geometric means in~\cite{Clayton:1996b}.
This generalized setting allows a straightforward treatment of fermions in a canonical setting (work in progress), where the primed constraints and algebra naturally appear.

\section*{Acknowledgements}

The author acknowledges support from the Natural Sciences and Engineering Research Council of Canada in the form of a postdoctoral fellowship.


\end{document}